\documentclass[pra,twocolumn]{revtex4-2}

	\usepackage{amsmath}
	\usepackage{bigints}
	\usepackage{mathtools}
	\usepackage{makeidx}
	\usepackage[normalem]{ulem}
	\usepackage{amsfonts}
	\usepackage{soul}
	\usepackage{color}
	\usepackage[latin1]{inputenc}
	\usepackage[usenames,dvipsnames]{pstricks}
	\usepackage{subfigure}
	\usepackage{epsfig}
	\usepackage{pst-grad} 
	\usepackage{pst-plot} 
	\usepackage[colorlinks,hyperindex]{hyperref}
	\usepackage{tikz,siunitx,mwe}%
	\hypersetup
	{
		colorlinks,%
		citecolor=black,%
		linkcolor=black,%
		urlcolor=black,%
	}

	\newcommand{\labeling}[8]{
}

\begin{document}
\title{Tunable $\Lambda$-type system made of a superconducting qubit pair }
 \par 

\author{Kuan-Hsun Chiang and Yung-Fu Chen\footnote{E-mail: yfuchen@ncu.edu.tw}}
\email{yfuchen@ncu.edu.tw}

\affiliation{Department of Physics, National Central University, Jhongli 32001, Taiwan}

\date{\today}

\begin{abstract}

Two transversely coupled and resonant qubits form symmetric and antisymmetric states as their eigenstates. In this paper, we show that parametric modulation of an individual qubit enables direct Rabi swapping between the two states. Its application to setup a $\Lambda$-type system with a pair of strongly coupled superconducting transmon qubits is discussed. The excited state is made of the symmetric state and the metastable state is the antisymmetric state. The coherence of the metastable state is only limited by the pure dephasing mechanism. Based on this scheme, $\Lambda$-type electromagnetically induced transparency, Autler-Townes splitting and stimulated Raman adiabatic passage are numerically demonstrated. We highlight the large frequency tunability in such superconducting $\Lambda$-type system.
\end{abstract}

\maketitle
\section{Introduction}
\label{sec:intro}
Quantum information processing and quantum optics with superconducting circuits have been experimentally realized in the past two decades~\cite{Blais04, Wallraff04, Koch07, Gu17, Krantz19, Arute19}. The semi-one-dimensional architecture of circuit quantum electrodynamics (QED) systems features large spatial mode matching among various types of artificial atoms and their detectors~\cite{Gu17}. The trade-off is that the eigenmodes of each quantum element have limited degrees of freedom, which thus restricts the diversity of selection rules. The lack of metastable states in these artificial atoms restricts the implementation of atomic and molecular optics in circuit QED systems. For instance, the category of $\Lambda$-type artificial atom has been rather unexplored. In most of the reported approaches, Purcell-protected qubits~\cite{Yang04, Liu16, Gu16, Novikov16, Long18}, additional coherent drive~\cite{Gu16, Long18}, and indirect Raman transition~\cite{Novikov16, Earnest18, Kelly10} are utilized. To set up a rather simple $\Lambda$-type artificial atom with an inherent metastable state, especially with decent frequency tunability, is still of great interest in superconducting quantum circuits.

An alternative strategy is to consider two coupled qubits, which provides an additional degree of freedom to the system, thereby modifying the parity characteristics of the eigenmodes. Symmetric and antisymmetric states of two coupled qubits are well known in literature~\cite{Dicke54, Ficek02}. Moreover, there are theoretical and experimental studies on coupled transmon qubits~\cite{KL13, vLoo13, Mlynek14, Majer07, Gambetta11, Srinivasan11, Zhang17}. The spontaneous decay of the symmetric (antisymmetric) state is inherently enhanced (suppressed), featuring its short (long) life time. Furthermore, the symmetric (antisymmetric) state is easy (difficult) to manipulate and measure. Nevertheless, the transition moment between these two extreme states is absent. It is known that two nonidentical Cooper-pair boxes form a system with nearly symmetric and antisymmetric states with weakly allowed transition between them~\cite{SR09}. Introducing e a mechanism that bridges symmetric and antisymmetric states could allow one to construct a promising $\Lambda$-type system. 

On the other hand, fast in-situ level tunability of superconducting qubits brings various conveniences to circuit QED systems. For example, manipulation of  qubit-qubit interaction through parametric modulation has been demonstrated~\cite{Niskanen07, McKay16, Roth17, Mundada19}. Here we introduce a simple parametric protocol to induce transition between symmetric and antisymmetric states in coupled identical qubits. It provides a tunable $\Lambda$-type scheme made of a resonant transmon qubit pair.

This paper is organized as follows. In Sec.~\ref{sec:ParametricDrive}, we introduce the protocol and the configuration of a general two-qubit system that constructs an effective $\Lambda$-type system. The results discussed in Sec~\ref{sec:ParametricDrive} are verified numerically by considering a pair of capacitively coupled transmon qubits in Sec~\ref{sec:lambda}. Sec.~\ref{sec:EIT} demonstrates its continuous-wave applications  on $\Lambda$-type electromagnetically induced transparency (EIT) and Autler-Townes splitting (ATS). Sec.~\ref{sec:STIRAP} demonstrates its pulsed control application on $\Lambda$-type stimulated Raman adiabatic passage (STIRAP). Sec.~\ref{sec:discussion} discusses the features and potential applications of this $\Lambda$-type system. Compared to other $\Lambda$-type systems, its large frequency tunability is also highlighted. Sec.~\ref{sec:conclusion} summarizes our paper.

\section{Parametric drive induced mode transfer}
\label{sec:ParametricDrive}
\begin{figure}
\includegraphics[width=85 mm]{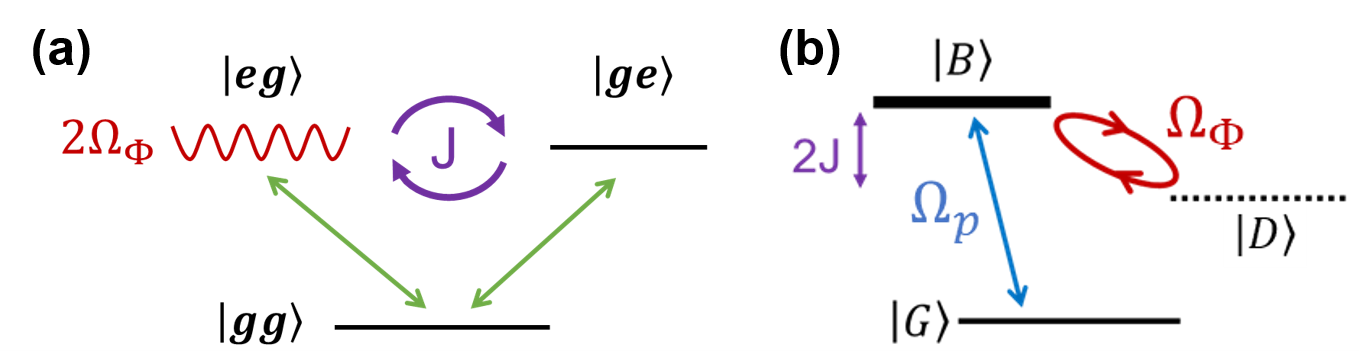}
\caption{
Level diagram of the tunable $\Lambda$-type system.  
(a) Representation of the control protocol in the basis of two resonant qubits $Q_a$ and $Q_b$. The transition frequency $\omega_a$ is modulated sinusoidally (red wavy, see main text), while $\omega_b$ stays static. $J$ represents the inter-qubit coupling. 
(b) Level diagram described in the system eigenstates, with corresponding transition driven by the parametric modulation in (a) (red circular). It is also referred to as the coupling beam in Sec.~\ref{sec:EIT} or the Stokes tone in Sec.~\ref{sec:STIRAP}. The blue arrow from $\vert G \rangle$ to $\vert B \rangle$ indicates the dipole-allowed transition, named as the probe beam in  Sec.~\ref{sec:EIT} or the pump tone in Sec.~\ref{sec:STIRAP}. 
} 
\label{fig:setup}
\end{figure}

Consider a pair of coupled qubits $Q_a$ and $Q_b$ (see Fig.~\ref{fig:setup}(a)). The Jaynes-Cummings Hamiltonian in the bare-qubit basis is written as

\begin{equation}
  \begin{aligned}
H_0 = \omega_a \vert eg \rangle \langle eg \vert + \omega_b \vert ge \rangle  \langle ge \vert + J [\vert eg \rangle \langle ge \vert + \vert ge \rangle \langle eg \vert],
  \end{aligned}
\label{eq:JC}
\end{equation}

\noindent where $\vert eg \rangle$ indicates that $Q_a$ is in the excited state and $Q_b$ is in the ground state. The excited state resembles dipole oscillation, which can be driven by an external electric field. $\omega_i$ is the transition frequency of $Q_i$. In the platform of superconducting qubits, the transverse inter-qubit coupling $J$ can be realized either by direct capacitive coupling or by an additional waveguide structure~\cite{Krantz19, Gu17}. A realistic example will be discussed in detail in Sec.~\ref{sec:lambda}. In the resonant case $\omega_a = \omega_b = \omega_0$, the three lowest eigenstates and eigenenergies of the combined system are (refer to Fig.~\ref{fig:setup}(b))

\begin{subequations}
\renewcommand{\theequation}{\theparentequation.\arabic{equation}}
  \begin{align}
\vert G \rangle & = \vert gg \rangle &, \quad \omega_G & = 0, \\
\vert D \rangle &= \frac{1}{\sqrt{2}}[\vert eg \rangle - \vert ge \rangle ] &, \quad \omega_D & = \omega_0 - J, \\
\vert B \rangle &=\frac{1}{\sqrt{2}}[\vert eg \rangle + \vert ge \rangle ] &,  \quad \omega_B & = \omega_0 + J.
  \end{align}
\label{eq:GDB}
\end{subequations}

 Consider that the spatial separation between the two qubits is much smaller than the transition wavelength $\lambda = 2 \pi c / \omega_0$, where $c$ is the speed of light. The two qubits see the oscillation of the electric field in phase. The antisymmetric state $\vert D \rangle$ resembles the out-of-phase dipole oscillation. It has zero total dipole moment and thus it is inherently protected from the external field and the vacuum fluctuation. Therefore, for the state $\vert D \rangle$ the spontaneous decay rate $\Gamma^1_D = 0$. It is also referred to as the \textit {Dark} state. On the other hand, the symmetric state  $\vert B \rangle$ features the in-phase dipole oscillation. Thus $\vert B \rangle$ has an enhanced interaction with the external field, with the total dipole moment twice that of the single qubit~\cite{Gambetta11, Srinivasan11}. Consequently, the spontaneous decay rate is enhanced by 4-fold, $\Gamma^1_B = 4\Gamma^1$. The $\vert B \rangle$ state is also referred to as the \textit{Bright} state~\cite{vLoo13, Mlynek14, Majer07, Gambetta11, Srinivasan11, Zhang17}. 

The decoherence rates of the state $\vert D \rangle$ and $\vert B \rangle$ are

\begin{subequations}
\renewcommand{\theequation}{\theparentequation.\arabic{equation}}
  \begin{align}
\gamma_{DD} &= \frac{1}{2}(\Gamma_a^{\phi}+\Gamma_b^{\phi}), \\
\gamma_{BB} &=  \frac{1}{2}(\Gamma_a^{\phi}+\Gamma_b^{\phi}) + \frac{1}{2}(4 \Gamma^1).
  \end{align}
\label{eq:decay}
\end{subequations}

\noindent Here $\Gamma_i^{\phi}$ is the pure dephasing rate of $Q_i$. Shining an electromagnetic wave is unable to induce transition between state $\vert D \rangle$ and $\vert B \rangle$ due to their parity characteristics (Fig.~\ref{fig:Selection}). However, it is reported that proper level modulation from Stark shift can induce coherent transfer between $\vert D \rangle $ and $\vert B \rangle$ ~\cite{Feng17}. Here we take advantage of the in-situ tunability of level spacing of a superconducting qubit. The transition frequency of $Q_a$ is sinusoidally modulated by its local flux line, denoted as $\omega_a(t) = \omega_0 + 2\Omega_{\Phi}(t) = \omega_0 +  2\Omega^0_{\Phi}(t)\sin{(\omega_{\Phi}t)}$. Here the factor 2 is introduced only for convenience. Meanwhile, $\omega_b = \omega_0 $ stays fixed (see Fig.~\ref{fig:setup}(a)). The Hamiltonian under parametric modulation reads

\begin{equation}
  \begin{aligned}
H_{\rm{m}}= [\omega_0 + 2\Omega_{\Phi}(t)] \vert eg \rangle  \langle eg \vert + \omega_0 \vert ge \rangle  \langle ge \vert \\ + [J \vert eg \rangle \langle ge \vert + \rm{H.c.}].
  \end{aligned}
\label{eq:bareHm}
\end{equation}

\noindent Rewriting $H_{\rm{m}}$ in the basis of $\{\vert G \rangle, \vert D \rangle, \vert B \rangle\}$, one gets

\begin{equation}
\begin{aligned}
H_{\rm{m }}= [ \omega_{D} + \Omega_{\Phi}(t) ] \vert D \rangle \langle D \vert + [\omega_B+ \Omega_{\Phi}(t) ] \vert B \rangle \langle B \vert \\
+  [\Omega_{\Phi}(t) \vert B \rangle \langle D \vert + \rm{H.c.}].
\label{eq:bareHmGDB}
\end{aligned}
\end{equation}

Eq.~\ref{eq:bareHmGDB} shows that the energy of the interested levels $\vert D \rangle$ and $\vert B \rangle$ vary in time in general. Nevertheless, for fast modulation $ \omega_{\Phi} \gg \Omega^0_{\Phi}$, motional averaging takes place~\cite{Shevchenko10, Li13, Silveri15, Wen20, LZS}, and one can ignore $ \Omega_{\Phi}(t)$ in the diagonal terms of Eq.~\ref{eq:bareHmGDB}. That is, the $\vert D \rangle$ and $\vert B \rangle$ levels act as if they stay at the centered frequencies:


\begin{equation}
\begin{aligned}
{\tilde{H}_{\rm{m }} = \omega_{D}  \vert D \rangle \langle D \vert+ \omega_B \vert B \rangle \langle B \vert+ [\Omega_{\Phi}(t) \vert B \rangle \langle D \vert + \rm{H.c.} ]}.
\label{eq:HmGDB}
\end{aligned}
\end{equation}

From the above discussion one easily sees that the parametric drive $2\Omega_{\Phi}(t) = 2\Omega^0_{\Phi}(t)\sin{(\omega_{\Phi}t)}$ is able to induce a coherent transition between state $\vert D \rangle$  and $\vert B \rangle$ as $\omega_{\Phi} \approx \omega_B - \omega_D = 2J$. When the effective Rabi frequency $\Omega^0_{\Phi}(t) \ll \omega_{\Phi}$, the rotating wave approximation (RWA) is satisfied. Therefore, $\omega_{\Phi} \approx 2J \gg \Omega^0_{\Phi}$ is the working regime of the scheme presented in this paper. The strong inter-qubit coupling is essential to achieve motional averaging and to avoid the breakdown of RWA.

In Eq.~(\ref{eq:bareHm}) to Eq.~(\ref{eq:HmGDB}), we consider the resonant condition, $\omega_a=\omega_b=\omega_0$. The parametric drive still works when the two qubits have unequal transition moment from their ground states to excited states. In such a case, the transition moment between the symmetric and antisymmetric states can be nonzero. However, the antisymmetric state is no longer ideally ''dark'', and so is the symmetric states~\cite{SR09}. Slight detuning between $\omega_a$ and $\omega_b$ has similar effect on the eigenstates, while the effective Rabi frequency $\Omega^0_{\Phi}(t)$ is slightly modified. See Appendix~\ref{sec:detune} for a related discussion. A coupled identical pair forms the exact bright state $\vert B \rangle$ and dark state $\vert D \rangle$. With the introduced parametric drive, direct Rabi swapping between the two extreme states become possible, without smearing their critical properties.


A mechanical analogy for the phenomenon discussed above is a pair of coupled identical oscillators with the vacuum mode $\vert G \rangle$, the out-of-phase oscillation $\vert D \rangle$ and the in-phase oscillation $\vert B \rangle$. By a sinusoidal modulation of one of the spring constants, the system picks up a relative phase between the two oscillators, resulting in the coherent mode transfer between the out-of-phase and the in-phase oscillation. The energy is injected or extracted by the external modulation agency. It is reported that similar principle is used to manipulate phonon modes in far-detuned coupled mechanical oscillators~\cite{Okamoto13}.

\section{Effective $\Lambda$-type system made of transmon pair}
\label{sec:lambda}

\begin{figure}
\includegraphics[width=85 mm]{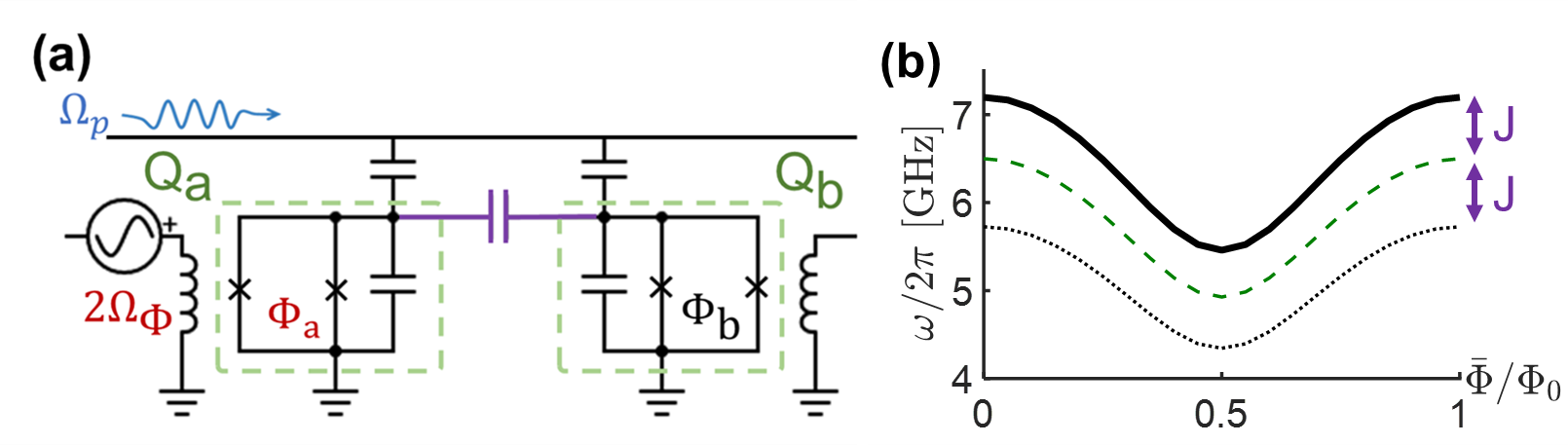}
\caption{Tunable $\Lambda$-type system is constructed with two transmon qubits.  
(a) Two capacitively coupled and resonant split-junction transmon qubits, with time-dependent flux $\Phi_a(t)$ threading the loop of $Q_a$. Level spacing of the on-resonance system is set by the static fluxes $\bar{\Phi}_a = \Phi_b =  \bar{\Phi}$ threading the two qubits. The parametric drive $\Omega_{\Phi}(t)$ is through the local flux line nearby $Q_a$. The electromagnetic excitation $\Omega_p$ is through the waveguide that capacitively coupled to $Q_a$ and $Q_b$. (b) Frequency tunability of the system. The transition frequency of the bright state $\omega_B$ (thick solid), the dark state $\omega_D$ (dotted) and bare qubit frequency $\omega_0$ (dashed green) as a function of dc flux bias $\bar{\Phi}$.} 
\label{fig:TCQ}
\end{figure}

The degenerate tunable coupling qubit (TCQ) architecture~\cite{Gambetta11, Srinivasan11, Zhang17}, i.e., two identical split junction transmon qubits with strong capacitive coupling, provides a good platform for our proposed $\Lambda$-type system. The schematic is shown in Figure~\ref{fig:TCQ}(a). The Hamiltonian of a TCQ in terms of the net Cooper pair number states $\{\vert n_i\rangle\}$ on qubit $i\in \{a, b\}$ reads
 
\begin{equation}
  \begin{aligned}
H_{\rm{TCQ}} &= \sum_{i=a,b}\sum_{n_i} 4E_{C_i}(n_{i}-n_{g_{i}})^2\vert n_i \rangle \langle n_i \vert \\
&- \sum_{i=a,b}\sum_{n_i}\frac{1}{2}E_{J_i}[\vert n_i+1 \rangle \langle n_i \vert + \vert n_i-1\rangle \langle n_i \vert] \\
&+ \sum_{i,j=a,b} \sum_{n_i, n_j} 4E_I n_i n_j \vert n_i \rangle \langle n_j \vert ,
  \end{aligned}
\label{eq:TCQcharge}
\end{equation}

\noindent where $n_{g_{i}}$ is the gate charge applied on $Q_i$. $E_{C_i}$ and $E_{J_i}$ denote the charging energy and Josephson energy of $Q_i$. $E_{C_a}=E_{C_b}=E_{C}$ is assumed. The inter-qubit coupling $J = 2E_I[E_{J_a}/E_{C_a}]^{1/4}[E_{J_b}/E_{C_b}]^{1/4}$~\cite{Gambetta11,Srinivasan11}, where $E_I$ denotes the interaction energy of the capacitively coupled two-qubit system. The TCQ is biased in degeneracy, i.e., the Josephson energy $\bar{E}_{J_a}(\Phi_a) = E_{J_b}(\Phi_b)$. We numerically solve the eigenstates of the TCQ in the charge basis $\{n_a,n_b\}$ from Eq.~(\ref{eq:TCQcharge}) (Appendix~\ref{sec:selection}). The characteristics of its lowest three eigenstates $\{ \vert G \rangle, \vert D \rangle, \vert B \rangle \}$ resemble that of Eq.~(\ref{eq:GDB}). As the parametric modulation is introduced,  $E_{J_a}(t) = E^{\rm{max}}_{J_a}\cos{[\pi\Phi_a(t)/\Phi_0]}\sqrt{1+d^2\tan^2{(\pi\Phi_a(t)/\Phi_0)}}$ varies in time, where $\Phi_0 =h/2e$ is the magnetic flux quantum and $d$ denotes the SQUID asymmetry~\cite{Koch07,Hutchings17}. Note that the selective modulation of $E_{J_i}(\Phi_i)$ is through the much stronger mutual inductance between the flux line and its nearest SQUID loop of $Q_i$, compared to that to the other qubit. It does not conflict the idea of the \textit {small} atom (molecule) assumption, which relies on small spatial separation between the antenna electrodes, compared to the wavelength of the field $\lambda$ in the case of transmon qubits. Refer to Appendix~\ref{sec:Crosstalk} for the discussion on the effect of finite flux crosstalk.

The dynamics of the lowest six eigenstates of the TCQ Hamiltonian (Eq. (7)), $\{ \vert G \rangle, \vert D \rangle, \vert B \rangle, \vert D2 \rangle, \vert E \rangle, \vert B2 \rangle \}$ is studied to examine the transition between $\vert D \rangle$ and $\vert B \rangle$ by a time-varing flux $\Phi_a(t)$. The density matrix $\rho$ written in the basis of these states is evolved via the master equation

\begin{figure}
\includegraphics[width=85mm]{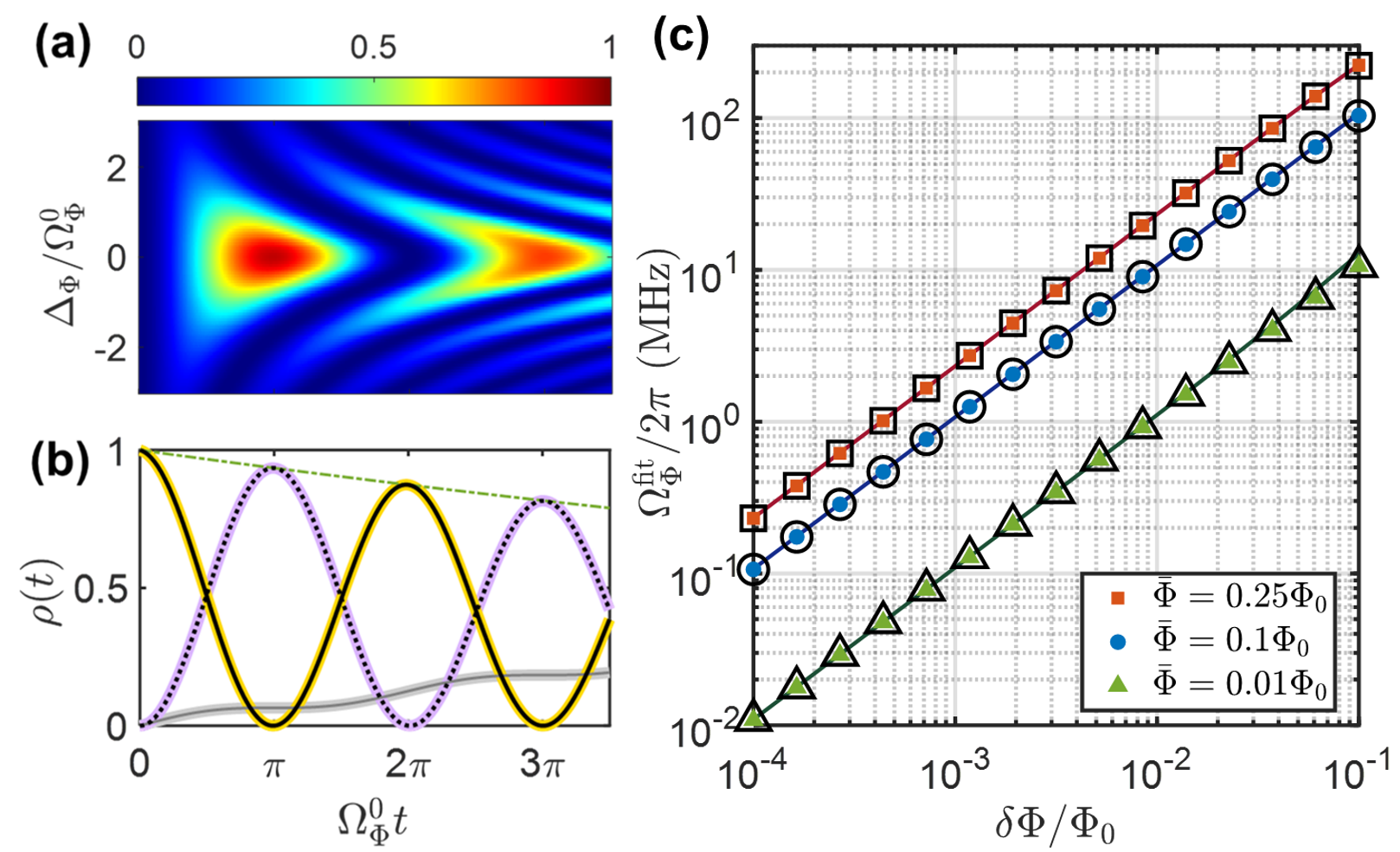}
\caption{
Numerical demonstration of parametrically induced transition between $\vert B \rangle$ to $\vert D \rangle$, started with $\rho(t=0) = \vert B\rangle \langle B\vert$. The external flux $\Phi_a(t) = \bar{\Phi} + \delta \Phi \cos{(\omega_{\Phi}t)}$ is applied to $Q_a$. $\bar{\Phi} = 0.25\Phi_0$. $\delta \Phi = 10^{-3}\Phi_0$, corresponds to $\Omega^0_{\Phi} /2\pi=$ 2.32 MHz. The parametric driving frequency $\omega_{\Phi}$ is set around $\omega_B - \omega_D \cong  2J $. 
(a) Rabi oscillation chevron. $\rho_{DD}$(colored) as a function of time, against detuning $\Delta_{\Phi}$. (b) Density matrix element $\rho_{DD}$(violet), $\rho_{BB}$(yellow) and $\rho_{GG}$(gray) as a function of time at $\Delta_{\Phi}=0$. The decay envelope (dashed) refers to the overall excitation decay rate $\Gamma_B^1/2$ to the ground state.
(c) Extracted Rabi frequency $\Omega^{\rm{fit}}_{\Phi}$ (hollow symbols) against flux modulation amplitude $\delta\Phi$ at different flux bias point $\bar{\Phi}$ with $\Delta_{\Phi}=0$. They agree well with the evaluation $\Omega^0_{\Phi} = \frac{1}{2}\frac{\partial \omega_a}{\partial \bar{\Phi}} \delta \Phi$ (solid lines). Refer to Appendix~\ref{sec:RWA} for $\rho(t)$ under different  $J/\Omega^0_{\Phi}$ ratio. The detailed mapping of $\Omega^{\rm{fit}}_{\Phi}$ and the validity of applying RWA against $\bar{\Phi}$ is in Appendix~\ref{sec:tunability}. Calculation results that includes only the lowest three levels of a TCQ are presented in (b) (black lines) and (c) (solid symbols) as well.
}
\label{fig:Lambda}
\end{figure}

\begin{equation}
\begin{aligned}
\dot{\rho}= &i [\rho, H_{\rm{TCQ}}] \\
 &+ \sum_{k,l}  {\bigl\{\frac{\Gamma_{jk}^1}{2} \mathcal{D}[\vert l\rangle \langle k \vert]\rho 
 +  \delta_{kl} \Gamma^{\phi} \mathcal{D}[\vert l \rangle \langle k \vert]\rho\bigr\}},
\label{eq:ME}
\end{aligned}
\end{equation}

\noindent where $\delta_{kl}$ is the Kronecker delta. $\vert k \rangle, \vert l \rangle \in \{ \vert G \rangle, \vert D \rangle, \vert B \rangle, \vert D2 \rangle, \vert E \rangle, \vert B2 \rangle \}$. Note that while truncation of higher energy state space is employed, the full Rabi Hamiltonian is considered in this six-state subspace simulation, i.e., the rotating wave approximation is not applied. The dissipation term $\mathcal{D}[\mathcal{O}]\rho = 2\mathcal{O}\rho\mathcal{O}^\dagger - \mathcal{O}^\dagger\mathcal{O}\rho - \rho\mathcal{O}^\dagger\mathcal{O}$. The term denotes that the spontaneous decay from $\vert B \rangle$ to $\vert D \rangle$ is absent due to the selection rule $\langle D \vert (  \hat{n}_a + \hat{n}_b )\vert B \rangle=0$.

For the following discussion, typical transmon parameters $E^{\rm{max}}_{J_a}/2\pi = E^{\rm{max}}_{J_b}/2\pi = E^{\rm{max}}_{J}/2\pi =15$ GHz, $E_{C}/2\pi = 400$ MHz and $E_I/2\pi=180$ MHz are used and the corresponding $J/2\pi = 700$ MHz. This strong inter-qubit coupling can be experimentally realized by direct capacitive coupling of 11.5 fF~\cite{Gambetta11, Srinivasan11, Zhang17}, and ensures that $\Omega^0_{\Phi}$ can achieve 100 MHz (see later discussion and Fig.~\ref{fig:Lambda}(c)). Note that, despite strong coupling $J$ being assumed here, the discussion based on the Jaynes-Cummings Hamiltonian in Sec. II is still valid. See Appendix~\ref{sec:FullRabiH} for the related discussion. The SQUID asymmetry factor $d=0.6$ is assumed.  The corresponding eigenfrequencies are $\omega_D/2\pi = 5.13$ GHz and $\omega_B /2\pi= 6.53$ GHz. As shown in Fig.~\ref{fig:TCQ}, the frequency of the bright state $\omega_B(\bar{\Phi})$ can be varied nearly 1 GHz by applying static flux biases. Potential applications of a tunable $\Lambda$-type system thus can be expected. As described by Eq.~(\ref{eq:decay}), the pure dephasing rate of the qubits $\Gamma_i^{\phi}$ limits the coherence of state $\vert D \rangle$. Throughout the discussion we set $\Gamma_i^{\phi}/2\pi = 0.2$ MHz, which is reported~\cite{Hutchings17} even if the transmons are biased away from their flux sweet spots. The spontaneous decay rate of the state $\vert B \rangle$, $\Gamma_B^1/2\pi =40$ MHz when embedded in a one-dimensional open transmission line and $\Gamma_B^1/2\pi =0.1$ MHz when embedded in a far-detuned resonator.

Figure~\ref{fig:Lambda} illustrates the parametric drive induced mode transfer by the aforementioned numerical method. By applying sinusoidal flux $\Phi_a(t) = \bar{\Phi} + \delta\Phi\sin(\omega_{\Phi}t)$ with $\omega_{\Phi} \approx  2J$, $\omega_a$ is parametrically modulated.  As a result, the atomic population can be swapped between $\vert B \rangle$ and $\vert D \rangle$ (Fig.~\ref{fig:Lambda}(a)). Spontaneous decay from $\vert B \rangle$ to the ground state $\vert G \rangle$ is present. In Fig.~\ref{fig:Lambda}(b), step-wise growing of $\rho_{GG}$ features the contrasting decay rate $\Gamma^1_{B}$ and $\Gamma^1_{D}$, respectively. The swapping rate $\Omega_{\rm eff} = \sqrt{{\Omega^0_{\Phi}}^2+\Delta^2_{\Phi}}$, where the Rabi frequency $\Omega^0_{\Phi} = \frac{1}{2}\frac{\partial \omega_a}{\partial \bar{\Phi}} \delta \Phi$. The detuning $\Delta_{\Phi}  \equiv \omega_{\Phi} - (\omega_B - \omega_D ) \cong \omega_{\Phi}  - 2J$.

The upper limit of the available $\Omega^0_{\Phi}$ depends on two factors : the flux bias point $\bar\Phi$ and the inter-qubit coupling $J$. Fig.~\ref{fig:Lambda}(c) demonstrates $\Omega^0_{\Phi}$ as a function of $\delta{\Phi}$ at different $\bar\Phi$, and therefore different $\omega_B$. Remarkably, the linear regime of $\Omega^0_{\Phi}$ versus $\delta{\Phi}$ covers a wide range from sub-MHz up to 100 MHz as long as $\Omega^0_{\Phi} \ll J$. The result shows a great potential for fast-control applications.

We also perform the simulation in the subspace constituted by only $\{ \vert G \rangle, \vert D \rangle, \vert B \rangle \}$ states. The three-state subspace simulation results are shown in Fig.~\ref{fig:Lambda}(b) (black lines) and (c) (solid symbols) for comparison. The effect of parametric drive induced mode transfer shows no difference between the three-state subspace and the six-state subspace simulations. It is because that the system has no nearby transition in response to the parametric drive except between $\vert D \rangle$ and $\vert B \rangle$ (see Appendix~\ref{sec:selection}). Meanwhile, the strong coupling $J$  has little effect on parametric drive induced transfer between $\vert D \rangle$ and $\vert B \rangle$ (see Appendix~\ref{sec:FullRabiH} for details). The appropriateness of the three-state subspace simulation echoes the principal idea of the parametrically induced transition based on the Jaynes-Cumming Hamiltonian (Eq.~(\ref{eq:JC})) described in Section~\ref{sec:ParametricDrive}. In the following sections, most of the numerical demonstrations are done in the three-state subspace to save computation power.

 \begin{figure*}
\includegraphics[width=170mm]{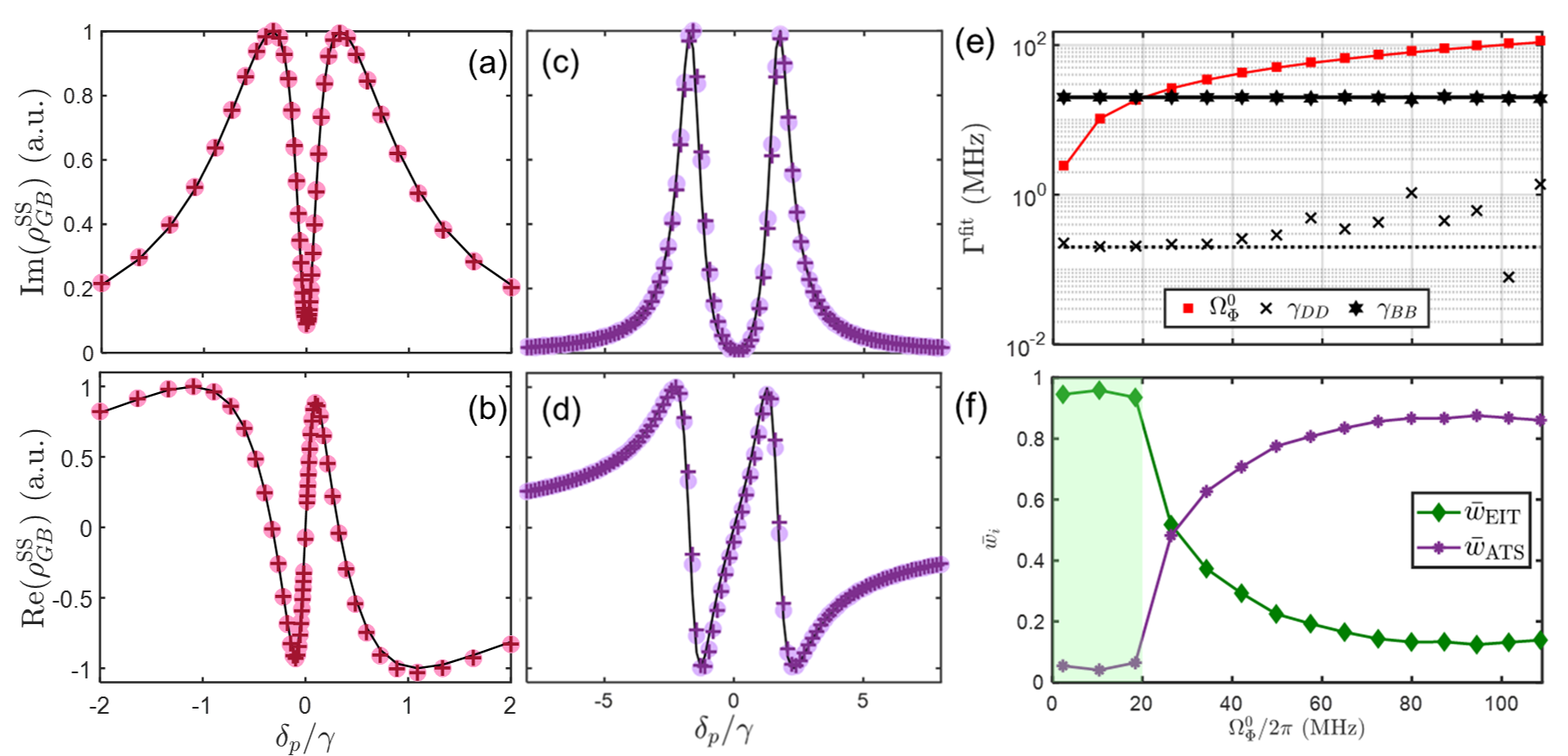}
\caption
{ 
Demonstration of $\Lambda$-type EIT and ATS in a degenerate TCQ. Refer to Fig.~\ref{fig:setup} for the setup and Sec.~\ref{sec:lambda} for the simulation parameters. Steady-state atomic absorption response Im($\rho^{SS}_{GB}$) as a function of probe detuning $\delta_p$ for (a) $\Omega^0_{\Phi}/2\pi = 13.1$ MHz and (c) $\Omega^0_{\Phi}/2\pi = 69.0$ MHz, respectively. (b) and (d) show the corresponding dispersion responses Re($\rho^{SS}_{GB}$) of (a) and (c), respectively. The symbols indicate the results of numerical simulation from Eq.~(\ref{eq:ME}). The results that include the 6-lowest levels of the system are also shown as crosses here for comparison. The fitting curve adapted from Eq.~(\ref{eq:GEN}) is presented as black solid lines. The crosses are appended to show the results where 6-lowest levels  are included. (e) Extracted fitting parameters $\gamma_{BB}/2\pi$ (black star), $\gamma_{DD}/2\pi$ (black cross) and $\Omega^0_{\Phi}/2\pi$(red square) of  Eq.~\ref{eq:GEN} to Im($\rho^{SS}_{GB}$) as a function of $\Omega^0_{\Phi}$. The lines indicate the parameter values used in the simulation. (f) $\bar{\omega}_{\rm{EIT}}$ and  $\bar{\omega}_{\rm{ATS}}$ against $\Omega^0_{\Phi}$. $\bar{\omega}_{\rm{EIT}} > 0.5$ indicates where the EIT spectrum fits better. The shaded region indicates the corresponding EIT regime $\Omega_{\Phi}^0 < \left|\gamma_{BB}-\gamma_{DD}\right|$.
}

\label{fig:EIT}
\end{figure*}

\section{Electromagnetically Induced Transparency and Autler-Townes Splitting}
\label{sec:EIT}

Consider a TCQ embedded in an open transmission line. As a probe beam $\Omega_p$ is shined onto the waveguide, together with an applied parametric drive $\Omega_{\Phi}$ onto the flux line (Fig.~\ref{fig:setup}(b)), the system Hamiltonian reads

\begin{equation}
\begin{aligned}
H_{\rm{m}}=&   \omega_{D} \vert D \rangle \langle D \vert  + \omega_{B} \vert B \rangle \langle B\vert\\
&+ [ \Omega_{p}(t) \vert G \rangle \langle B \vert + \Omega_{\Phi}(t) \vert D \rangle \langle B\vert + \rm{H.c.}. ]
\label{eq:HLambda}
\end{aligned}
\end{equation}

\noindent The weak probe beam $\Omega_{p}(t) = \Omega_{p}^0\sin{(\omega_pt)}$ with detuning $\delta_p \equiv \omega_p - \omega_{B}$. The coupling beam $\Omega_{\Phi}(t)=\Omega_{\Phi}^0\sin{(\omega_{\Phi}t)}$ with detuning $\Delta_{\Phi} = 0$. In Fig.~\ref{fig:EIT}, the master equation (Eq.~\ref{eq:ME}) is evolved until a steady-state is reached to spectroscopically demonstrate $\Lambda$-type EIT and ATS in a one-dimensional open transmission line, where we set $\Gamma_B^1/2\pi=40$ MHz and $\Gamma_i^{\phi}/2\pi = 0.2$ MHz. Correspondingly $\gamma_{BB}/2\pi=40.2$ MHz and $\gamma_{DD}/2\pi=0.2$ MHz. Ref.~\cite{Sun14} gives the general solution of the steady-state optical susceptibility $\chi_{GB}(\delta_p)$ of Eq.~(\ref{eq:HLambda})

\begin{equation}
\begin{aligned}
\chi_{GB}(\delta_p) =  \frac{\vert p_{GB}\vert^2}{\delta_- - \delta_+} [\frac{\delta_++i\gamma_{DD}}{\delta_p-\delta_+} -\frac{\delta_-+i\gamma_{DD}}{\delta_p-\delta_-}] ,
\label{eq:GEN}
\end{aligned}
\end{equation}

\noindent where $\delta_{\pm} = (-i\gamma_{DD}-i\gamma_{BB}\pm\Omega_T)/2$, with $\Omega_T = \sqrt{{\Omega^0_{\Phi}}^2 - (\gamma_{DD} - \gamma_{BB})^2}$. $p_{GB} = \langle G \vert ( \hat{n}^2_a + \hat{n}^2_b ) \vert B \rangle$ denotes the transition moment between $\vert G \rangle$ and $\vert B \rangle$ . Additionally, $\chi_{GB}(\omega_p)= \tilde{\alpha}\rho^{SS}_{GB}(\omega_p)$. $\rho^{\rm{SS}}_{GB}(\omega_p)$ is the steady-state oscillation amplitude of the density matrix element. The factor $\tilde{\alpha} = \Omega_p^0/(\varepsilon_0\vert E_p \vert^2 )$, with  $\varepsilon_0$ the vacuum permitivity and $E_p$ the amplitude of the probe field.

The coherence criteria of $\Lambda$-type EIT, $\gamma_{BB} \gg \gamma_{DD}$, is satisfied in this system. The EIT effect competes with the ATS one. When $\Omega_{\Phi}^0 < \left|\gamma_{DD}-\gamma_{BB}\right|$, resembling that the control tone Rabi frequency is smaller than the linewidth of $\vert B \rangle$ state, quantum interference dominates. Therefore, Eq.~(\ref{eq:GEN}) reduces to $\chi_{\rm{EIT}}$, and a narrow Lorenztian transmission window emerges at the center of $\vert G \rangle$ to $\vert B \rangle$ absorption, as illustrated in Fig.~\ref{fig:EIT}(a)(b). As $\Omega_{\Phi}^0 > \left|\gamma_{DD}-\gamma_{BB}\right|$, ATS~\cite{Silla09, Jian11} takes over the optical response, Eq.~(\ref{eq:GEN}) reduces to $\chi_{\rm{ATS}}$, two Lorentzian absorption window separated by $\Omega^0_{\Phi}$, as illustrated in Fig.~\ref{fig:EIT}(c)(d). In both regimes, our simulation based on Eq.~(\ref{eq:ME}) matches the analytical solution Eq.~(\ref{eq:GEN}) with excellent agreement. Fig~\ref{fig:EIT}(e) shows the fitting parameters $\{\gamma_{DD}, \gamma_{BB}, \Omega^0_{\Phi}\}$ of the simulated Eq.~(\ref{eq:GEN}) to ${\rm{Im}}(\rho^{\rm{SS}}_{GB})$ for different modulation amplitude $\Omega^0_\Phi$. The result implies the validity of the effective $\Lambda$-type system activated by parametric drive induced transition.

  Akaike's information metric~\cite{Anisimov11} is performed to analyze the steady-state optical response ${\rm{Im}}(\rho^{\rm{SS}}_{GB})$ as a function of $\Omega^0_{\Phi}$ (Fig.\ref{fig:EIT}(f)). ${\rm{Im}}(\rho^{\rm{SS}}_{GB})$ are fitted to both $\chi_{\rm{EIT}}$ and $\chi_{\rm{ATS}}$ respectively. The per-point weight 
  
\begin{subequations}
\renewcommand{\theequation}{\theparentequation.\arabic{equation}}
\begin{align}
  \bar{w}_{\rm{EIT}} &=  \frac{e^{-\bar{I}_{\rm{EIT}}}} { e^{-\bar{I}_{\rm{EIT}}} + e^{-\bar{I}_{\rm{ATS}}}} \\
  \bar{w}_{\rm{ATS}} &= 1 - \bar{w}_{\rm{EIT}}
\label{eq:AIC}
\end{align}
\end{subequations}
  
\noindent are then evaluated, where $\bar{I} = \ln{R/N} + 2k/N$, with $N$ the number of data points, $R$ the sum of the square of fitting residual and $k$ the number of fitting parameters~\cite{Liu16}.  Ref.~\cite{Sun14} gives the EIT window of the control tone power, $2\gamma_{DD}\sqrt{\gamma_{DD}/(\gamma_{BB}+2\gamma_{DD})}<\Omega^0_{\Phi}< \left|\gamma_{DD}-\gamma_{BB}\right|$, which gives 0.04 MHz$<\Omega^0_{\Phi}/2\pi<$ 20 MHz in our scheme. It agrees with the measure in Fig.~\ref{fig:EIT}(f).

The non-unity transparency at the center is limited by the finite decoherence rate $\gamma_{DD}$ of the metastable state $\vert D \rangle$. Practically, 1 MHz $ < \gamma_{BB} /2\pi < $ 50 MHz is dominated by the spontaneous decay, while 0.1 MHz $ < \gamma_{DD} /2\pi < $ 10 MHz is dominated by the pure dephasing. Consequently, the bottleneck threshold for the emergence of EIT, $\gamma_{BB} > 2 \gamma_{DD}$, can be easily overcome. Here we illustrate the case $\gamma_{BB} / \gamma_{DD} \cong 100$. Remark that in principle $\gamma_{BB}$ can be enhanced while $\gamma_{DD}$ can be suppressed independently, owing to their contrary decoherence channels.

\section{Stimulated Raman adiabatic passage}
\label{sec:STIRAP}
 \begin{figure}
\includegraphics[width=75mm]{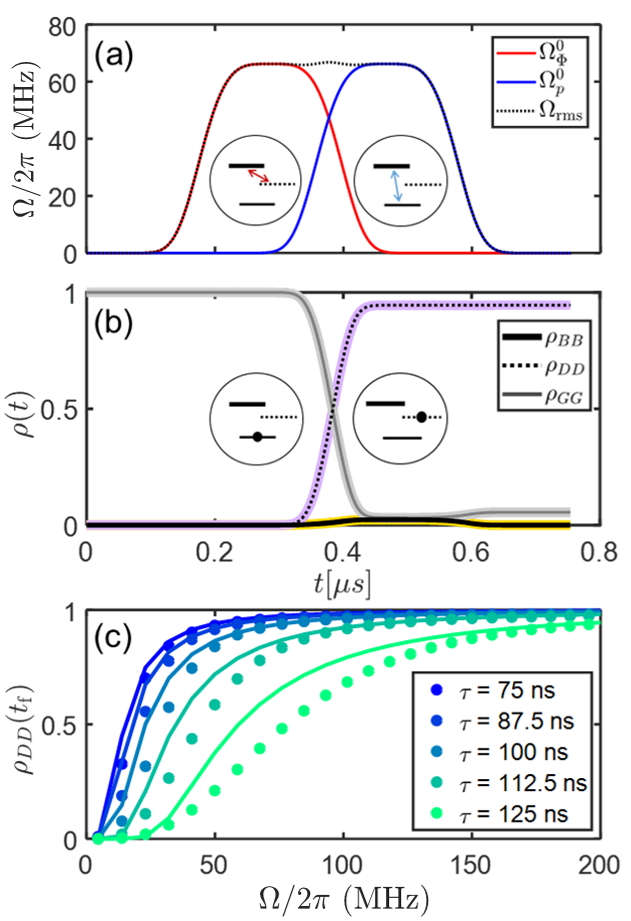}
\caption{
$\Lambda$-type STIRAP to swap population from $\vert G \rangle$ to $\vert D \rangle$ state, started with $\rho({t=0}) = \vert G \rangle \langle G \vert$ and stopped at time $t_{\rm{f}}=0.8 \rm{\mu}$s $= 2.57\times 2\pi/\gamma_{BB}$.
Refer Fig.~\ref{fig:setup}(b) for the the setup. 
(a) Control sequence, with the parametric Stokes pulse $\Omega^0_{\Phi}(t)$ (red) followed by the pump pulse $\Omega^0_p(t)$ (blue). The pulse separation $2\tau = 182$ ns and the pulse duration $2T = 200$ ns. The black dotted line indicates $\Omega^0_{\rm{rms}}(t)$ with peak Rabi frequency 67.2 MHz.
(b) Corresponding atomic response as a function of time. $\rho_{DD}(t_{\rm{f}}) = 0.969$ (dotted) and $\rho_{GG}(t_{\rm{f}}) = 0.031$ (gray solid). The intermediated state $\rho_{BB}$ (thick solid black) reaches 0.014 during the process. Simulation results that include the 6-lowest levels of a TCQ are presented as colored shaded lines for comparison.
(c) Transfer effeciency as a function of peak pulse amplitude $\Omega^{\rm{pk}}$. Colors represent different pulse delay $\tau$. Solid lines represent the analytical modeling accounting for nonadiabaticity~\cite{Yatsenko02, Yatsenko14}.
}
\label{fig:STIRAP}
\end{figure}

In this section, we show the application of proposed $\Lambda$-type system in transient optical response by considering STIRAP \cite{ Vitanov17, Torosov13, Bergmann19, Kumar16,AV19, Premaratne17}. Refer to Eq.~(\ref{eq:HLambda}) for the system Hamiltonian. The Stokes tone $\Omega_{\Phi}(t) = \Omega^0_{\Phi}(t)\sin(\omega_{\Phi}t)$ is parametrically applied while the pump tone $\Omega_p(t)= \Omega^0_p(t)\sin(\omega_pt)$ is applied onto the waveguide. A typical STIRAP sequence is composed of two partially overlapped pulses in time domain,  $\Omega_{\Phi} (t)$ followed by  $\Omega_p (t)$, with equal peak Rabi frequency $\Omega^{\rm{pk}}$. With an optimized pulse sequence, the population can be transferred from $\vert G \rangle$ to $\vert D \rangle$ with negligible population in the dissipative intermediate state $\vert B \rangle$. Fig.~\ref{fig:STIRAP}(a) shows an illustrative protocol of a $\Lambda$-type STIRAP and Fig.~\ref{fig:STIRAP}(b) is the simulated corresponding response of the system, where the master equation (Eq.~\ref{eq:ME}) is used. The detunings $\delta_p = \Delta_{\Phi} = 0$ are assumed. The atomic state is swapped from $\vert G \rangle$ to $\vert D \rangle$ by a single set of hyper-Gaussian pulse, $\Omega^0_p(t) = \Omega^{\rm{pk}}\rm{exp}[-(t-\tau/2)^4/T^4] $ and $\Omega^0_{\Phi}(t) = \Omega^{\rm{pk}}\rm{exp}[-(t+\tau/2)^4/T^4]$. Here $\tau$ is the pulse separation and $2T$ characterizes the pulse width. The transfer efficiency is defined as the metastable state population at the end of the process,  $\rho_{DD}(t_{\rm{f}})$.  The non-unit transfer efficiency of the process is contributed by the nonadiabaticity of the protocol, and described by~\cite{Yatsenko02, Yatsenko14}
 
\begin{equation}
\rho_{BB}(t_{\rm{f}}) = \exp{ \biggl\{ - \bigintsss_0^{t_{\rm{f}}}{4\gamma_{BB} {\biggl(\frac{\dot{\theta}}{\Omega^0_{\rm{rms}}}\biggr)}^2\,dt} \biggr\}},
\label{eq:STIRAPadb}
\end{equation}

\noindent where $\Omega^0_{\rm{rms}}(t) = \sqrt{{\Omega^0_{\Phi}}(t)^2 + \Omega^0_p(t)^2} $ is the average Rabi frequency of the two pulses and $\theta(t) = \tan^{-1}{[\Omega^0_p(t) / \Omega^0_{\Phi}(t)]}$ is the mixing angle. $\rho_{DD}(t_{\rm{f}})$ approaches unity as the local adiabatic condition for the sequential pulses $\Omega^0_{\rm{rms}}(t) \gg \lvert{\dot{\theta}(t)}\rvert$ is satisfied.

Fig.~\ref{fig:STIRAP}(c) shows that the incomplete population transfer $\rho_{DD}(t_{\rm{f}})$ with various protocol parameters \{$\tau$, $\Omega$\} can be explained by Eq.~(\ref{eq:STIRAPadb}). Remark that despite the length of the sequence comparable to the lifetime of the intermediate state $\vert B \rangle$, the transfer efficiency can still approach unity as long as adiabaticity is hold throughout the process.

There are pioneer experiments that demonstrate the STIRAP process in a $\Xi$-type artificial atom or a qubit-cavity system in circuit QED architecture \cite{KKumar16, Premaratne17, AV19}. Here we address the possibility of applying STIRAP on this effective $\Lambda$-type artificial atom, with a relaxation-free final state. Note that despite large excited state decoherence $\gamma_{BB}$, high transfer efficiency $\rho_{BB}(t_{\rm{f}})$ is obtained. With the TCQ being placed in a far-detuned cavity, the system with suppressed $\gamma_{BB}$ could approach unit transfer efficiency.

In the sense of coupled oscillator dynamics, this STIRAP process is analogous to the cooperation of the in-phase drive on both objects and stiffness modulation on one of them. It excites the system from vacuum to out-of phase oscillation, without any in-phase oscillation.

\section{Discussion}
\label{sec:discussion}

Realization of EIT in superconducting quantum circuits has been studied and carried out in recent years~\cite{AS10, Anisimov11, Sun14, Gu16, Liu16, Novikov16, Long18,Andersson20}. The main challenge is to create a reliable metastable state~\cite{AS10, Anisimov11, Sun14}. In most of the successful approaches to create effective $\Lambda$-type levels, a single qubit is dispersively coupled with a cavity (QC). A strong pumping field is also applied either to excite a higher order transition~\cite{Novikov16} or to modify the level configuration~\cite{Gu16, Long18}. They overcome the coherence criteria by making use of the decay rates of the cavity-like state (excited state) and the Purcell-protected qubit-like state (metastable state). However, the desires to have a high cavity decay rate $\kappa$ and to have low Purcell-protected qubit decay $\gamma_{\kappa} = (g/\Delta)^2\kappa$ conflict each other, thereby compromising the choice of the cavity bandwidth.  Here $g$ and $\Delta$ denote the coupling and detuning between the qubit and the cavity, respectively.  Regarding frequency tunability of the QC approach, a tunable qubit can slightly modify the transparency window by $g^2/\Delta$ as it moves toward the cavity frequency. Meanwhile, degraded Purcell protection limits the performance of the metastable state in the QC approach. Alternatively, SQUID loops with high critical current can be integrated to the coplanar waveguide cavity to achieve high tunability, while the manufacturing would become complicated. In contrast, the degenerate TCQ itself is a $\Lambda$-type system, with ideally zero spontaneous decay from the Dark state. It allows a tunable transparency window up to few GHz (Fig.\ref{fig:TCQ}(b)) with standard SQUID loops on both qubits. Meanwhile, the spontaneous decay remain eliminated~\cite{Gambetta11, Srinivasan11} and insignificant degradation of the pure dephasing could be achieved~\cite{Hutchings17}. Fig.~\ref{fig:EIT}(b) measures the slow light effect with $v_g\cong 5\% $ of the speed of light. Benefit from the tunability of the system, frequency transduction between the encoded and retrieved light could be possible in such a quantum memory.

Note that the main difference between this scheme and a conventional $\Lambda$-type atomic level is that a negligible spontaneous decay from $\vert B \rangle$ to $\vert D \rangle$ is mediated by the local flux lines with noise near $\omega_B - \omega_D$. $\Gamma^1_{DB}$ could be below 1 Hz~\cite{Koch07, Hutchings17}. Therefore, frequency down-conversion due to spontaneous emission~\cite{Koshino13, Inomata14} is absent in this system.

The proposed scheme can be generalized for different platforms. For example, it can be applied to different types of qubits, as well as with alternative realization of strong transverse coupling. The shortcoming of this scheme is that the frequency tunability of the system introduces non-negligible dephasing, which is the main decoherence source of metastable state. Advanced effort can be made to suppress the pure dephasing either by surface treatment~\cite{Kumar16} or considering different types of qubits~\cite{Yan16}.

\section{Conclusion}
\label{sec:conclusion}
In conclusion, we propose a simple and effective $\Lambda$-type system made of a resonant superconducting qubit pair. The symmetric state plays the role of the excited state and the antisymmetric state represents the metastable state. They are mediated by a parametric drive on one of the qubits. Its application on $\Lambda$-type EIT, ATS and STIRAP are numerically demonstrated. Compared to other approaches in circuit QED architecture, our proposed $\Lambda$-type scheme features large level tunability, while retaining sufficient coherence of the metastable state. The device volume is compact and the manufacturing can be directly implemented by the transmon-type approach. It provides a solution of having a $\Lambda$-type system in on-chip superconducting quantum circuits.

\begin{acknowledgments}

The authors wish to thank Io-Chun Hoi, Jeng-Chung Chen and George Thomas for helpful discussion. This work has been supported by Ministry of Science and Technology in Taiwan under Grants No. MOST 109-2112-M-008-025 and MOST-110-2112-M-008-024.

\end{acknowledgments}

\appendix
{

\section{PARAMETRIC MODULATION ON DETUNED QUBITS} 
\label{sec:detune}

Consider detuned qubits with modulation on $Q_a$

\begin{equation}
H_{\rm m} = 
\begin{bmatrix}

\omega_a+2\Omega_{\Phi} & J\\
J & \omega_b\\

\end{bmatrix}
\rm{in \ the \ basis \ of} 
\begin{pmatrix}
\vert eg \rangle \\
\vert ge \rangle
\end{pmatrix},
\label{eq:DetunedBQB}
\end{equation}

\noindent where $\omega_{a,b} = \omega_0 \pm \delta$. The two lowest excited states of the system read

\begin{subequations}
\label{eq:DetunedEigenstates}

\begin{align}
\vert - \rangle &= \frac{1}{\sqrt{2G(G+\delta)}}  [ (-J \vert eg \rangle + (G+\delta) \vert ge \rangle ],
\label{eq:DetunedEigenstates1}\\
\vert + \rangle &= \frac{1}{\sqrt{2G(G+\delta)}} [ (G+\delta) \vert eg \rangle + J \vert ge \rangle ]\label{eq:DetunedEigenstates2}
\end{align}

\end{subequations}

\noindent where $G \equiv \sqrt{J^2+\delta^2}$. The Hamiltonian Eq.~(\ref{eq:DetunedBQB}) described in the basis of Eq.~(\ref{eq:DetunedEigenstates}) becomes

\begin{equation}
H_{\rm m} =  
\begin{bmatrix}
{\omega_0-G+\Omega_{\Phi}(1-\frac{\delta}{G})} & {-\Omega_{\Phi}J/G} \\
{-\Omega_{\Phi}J/G} & {\omega_0+G+\Omega_{\Phi}(1-\frac{\delta}{G})}
\label{eq:HmDetune}
\end{bmatrix}.
\end{equation}

Compared to Eq.~(\ref{eq:HmGDB}) which describes the zero-detuning, the off-diagonal terms in Eq.~(\ref{eq:HmDetune}) are scaled by $J/G\leq1$. As a result, depsite with finite detuning, the parametric drive activates significant swapping between the lowest two eigenstates of the system as long as $J \gg \delta$. One may realize that the moment to parametric drive is proporitonal to $\frac{\partial\omega}{\partial\Phi}\frac{J}{G}$. As the TCQ is in degeneracy, the $\vert - \rangle$ state is decay-free and has forbidden transition to the $\vert + \rangle$ state, the parametric drive is of particular interest. 

\section{EIGENSTATES OF A DEGENERATE TUNABLE COUPLING QUBIT} 
\label{sec:selection}
\begin{figure}
\includegraphics[width=85mm]{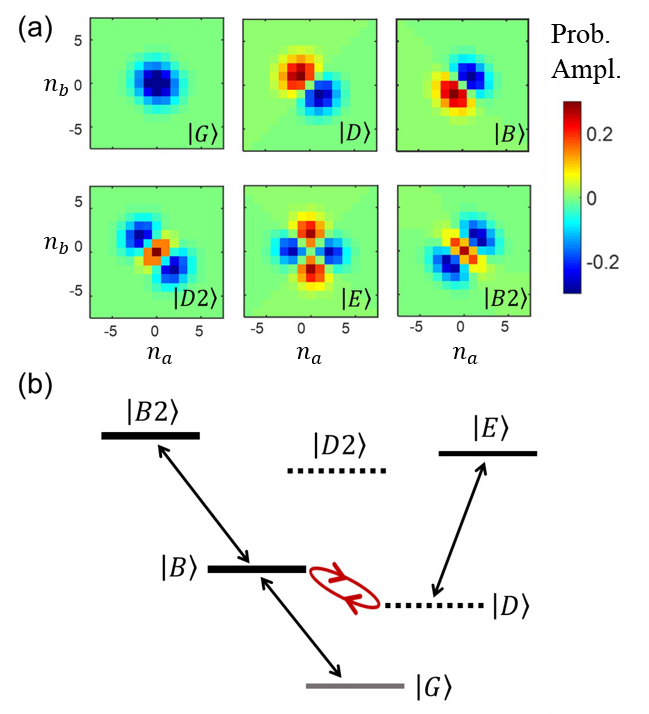}
\caption{
Energy levels and selection rules of a degenerate TCQ.
(a) Six lowest eigenstates $\{ \vert j \rangle \}$ of a degenerate TCQ. They are presented in the basis of \{$n_a, n_b$\}, with truncated net cooper pair number $n_i \in \{ -7, 7\}$. The color bar represents the probability amplitude $\psi(n_a,n_b)$. The opposite sign (red and blue) only indicates the relative phase.
(b) The selection rules between state $\vert j \rangle$ and $\vert k \rangle$, determined by the transition moment $\langle j \vert \hat{N} \vert k \rangle$, with $\hat{N}$ the number operator of Cooper pairs.  The allowed transition (black arrows) and parametrically available driving channel (red circular arrow) are indicated.}
\label{fig:Selection}
\end{figure} 

The six lowest eigenstates of a degenerate TCQ in the $\{n_a, n_b\}$ basis are presented in Fig.~\ref{fig:Selection}. Consider an identical transmon pair. $\{\vert e \rangle, \vert f \rangle \}$ denote their first excited state and second excited states, respectively. The upper three eigenstates are $\vert E \rangle  = \vert ee \rangle $ , $\vert D2 \rangle = [\vert fg \rangle - \vert gf \rangle ] / \sqrt{2}$ and $\vert B2 \rangle = [\vert fg \rangle + \vert gf \rangle ] / \sqrt{2}$, respectively. In addition to the $\Lambda$-type level structure, one can access $N$-type transition with state $\vert E \rangle$ involved (Fig.~\ref{fig:Selection}(b))~\cite{SR09}.  Note that there is no similar parametric driving channel between $\vert D2 \rangle$ and $\vert B2 \rangle$ since transverse coupling is absent between $\vert fg \rangle$ and $\vert gf \rangle$.

\section{EFFECT OF FLUX CROSSTALK} 
\label{sec:Crosstalk}

\begin{figure}[!t]
\includegraphics[width=85mm]{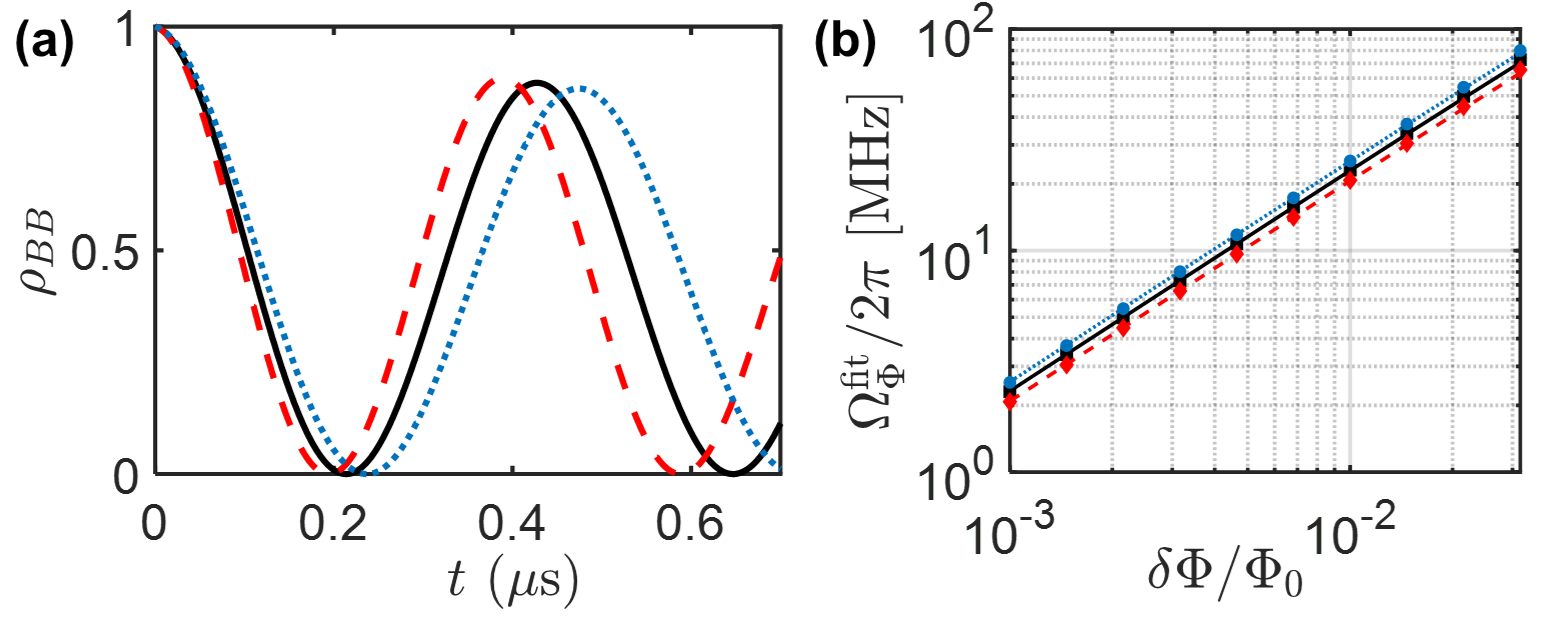}
\caption
{
Simulation of parametric drive with an overestimated flux crosstalk $C_{\rm{f}}=0.1$ between $Q_a$ and $Q_b$.
(a) Effective Rabi oscillation of $\rho_{BB}(t)$. The black line indicates $C_{\rm{f}}=0$. The crosstalk value with $\Omega^b_{\Phi}=0.1\Omega^a_{\Phi}$(red dashed) and $\Omega^b_{\Phi}=-0.1\Omega^a_{\Phi}$(blue dotted) are presented as well.
(b) Fitted Rabi frequency $\Omega^{\rm{fit}}_{\Phi}$ as a function of parametric driving amplitude $\delta\Phi_a$ on $Q_a$. The symbols represent the simulation results and the lines gives the deduction from Eq.~(\ref{eq:dOmg_cstk}). The setting parameters here are identical to that of Fig.~\ref{fig:Lambda}.
}

\label{fig:Crosstalk}
\end{figure} 

Neighboring transmon qubits could suffer from residual flux crosstalk. Typically it could as low as 1\%~\cite{Zhong21}. To formalize the problem,  one considers modulation on both tunable qubits,

\begin{equation}
  \begin{aligned}
H_{\rm{m}} = [\omega_0 + 2\Omega^a_{\Phi}(t)] \vert eg \rangle + [\omega_0 + 2\Omega^b_{\Phi}(t)] \vert ge \rangle \\
+ J [\vert eg \rangle \langle ge \vert + \vert ge \rangle \langle eg \vert].
  \end{aligned}
\label{eq:bareH}
\end{equation}

\noindent The Hamiltonian expressed in the basis of \{$\vert G \rangle$, $\vert D \rangle$, $\vert B\rangle$\}  can be written as
\begin{equation}
\begin{aligned}
H_{\rm{m}}  = [ \omega_{D} +\bar\Omega(t) ] \vert D \rangle + [ \omega_{B} +\bar\Omega(t) ] \vert B \rangle +\\  \delta\Omega(t) [\vert B \rangle \langle D \vert + \vert D \rangle \langle B \vert],
\label{eq:equation3}
\end{aligned}
\end{equation}

\noindent where 

\begin{subequations}
\renewcommand{\theequation}{\theparentequation.\arabic{equation}}
\begin{align}
\bar\Omega(t) &= \Omega^a_{\Phi}(t) + \Omega^b_{\Phi}(t)\\
\delta\Omega(t) &= \Omega^a_{\Phi}(t) - \Omega^b_{\Phi}(t).
\label{eq:dOmg}
\end{align}
\end{subequations}

Consider nonzero $\Omega^b_{\Phi}(t)$ comes from $\delta\Phi_a$ due to finite flux crosstalk,  Eq.~(\ref{eq:dOmg}) can be written as

\begin{equation}
\begin{aligned}
\delta\Omega(t) &= \Omega^a_{\Phi}(t) - \frac{1}{2}\frac{\partial{\omega_b}}{\partial{\Phi_b}}C_{\rm{f}}\delta\Phi_a
\label{eq:dOmg_cstk}
\end{aligned}
\end{equation}

\noindent where the factor $0<C_{\rm{f}}<1$ denotes the level of flux crosstalk. Note that in general the sign of $\frac{\partial \omega_b}{\partial \bar{\Phi}}$ can be either positive or negative, depending on the flux bias point. Therefore, the effective Rabi frequency is either enhanced or suppressed by the factor $C_{\rm{f}}$. Fig.~\ref{fig:Crosstalk} shows the modified effective Rabi frequency with an over-estimated cross talk ratio $C_{\rm{f}}=0.1$. Here we demonstrate that with finite flux crosstalk to the neighboring qubit, the effect of parametric drive remains the same, while the magnitude is slightly modified.

\section{FULL RABI HAMILTONIAN} 
\label{sec:FullRabiH}

Consider a pair of resonant qubits $\omega_a=\omega_b=\omega_0$ with strong transverse coupling $J$, the full Rabi Hamiltonian in the bare-qubit bases $\{ \vert gg \rangle, \vert eg \rangle, \vert ge \rangle, \vert ee \rangle\}$ reads

\begin{equation}
H^R_{\rm 0} = 
\begin{bmatrix}
0 & 0                                & 0 & J\\
0 & \omega_0 & J & 0\\
0 & J & \omega_0         &  0\\
J & 0 & 0        &  2\omega_0

\end{bmatrix}.
\end{equation}

\noindent The eigenvectors are
\begin{widetext}
\begin{equation}
\vert G \rangle = 
\frac{1}{\sqrt{2\Delta(\Delta+\omega_0)}}
\begin{pmatrix}
-\omega_0-\Delta \\0 \\0 \\J
\end{pmatrix}
,
\vert D \rangle = 
\frac{1}{\sqrt{2}}
\begin{pmatrix}
0\\1 \\-1 \\0
\end{pmatrix}
,
\vert B \rangle = 
\frac{1}{\sqrt{2}}
\begin{pmatrix}
0 \\1 \\1  \\0
\end{pmatrix}
,
\vert E \rangle = 
\frac{1}{\sqrt{2\Delta(\Delta+\omega_0)}}
\begin{pmatrix}
J \\0 \\0 \\ \omega_0+\Delta
\end{pmatrix},
\end{equation}
\end{widetext}
where $\Delta \equiv \sqrt{\omega_0^2+J^2}$. When the parametric modulation on $Q_a$ is introduced, i.e., $\omega_a(t)=\omega_0+2\Omega_{\Phi}(t)$, the full Rabi Hamiltonian in the eigenbases $\{ \vert G \rangle, \vert D \rangle, \vert B \rangle, \vert E \rangle\}$ becomes

\begin{widetext}
\begin{equation}
H^R_{\rm m} = 
\begin{bmatrix}
{(\omega_0-\Delta)(1-\frac{\Omega_{\Phi}(t)}{\Delta})} & {0}   & {0} & { \frac{J}{\Delta} \Omega_{\Phi}(t) }\\
{0} & {\omega_0-J+\Omega_{\Phi}(t)} & { \Omega_{\Phi}(t) } & {0} \\
{0} & { \Omega_{\Phi}(t) } & {\omega_0+J+\Omega_{\Phi}(t) }  &  {0} \\
{\frac{J}{\Delta} \Omega_{\Phi}(t) } & {0} & {0}        &{(\omega_0+\Delta)(1+\frac{\Omega_{\Phi}(t)}{\Delta})}\\ 
\end{bmatrix}.
\label{eq:HmRabi}
\end{equation}
\end{widetext}

From Eq.~(\ref{eq:HmRabi}) one realizes that  : (1) The effects of parametric modulation $\Omega_{\Phi}$ on the $\{\vert G\rangle$, $\vert D\rangle$, $\vert B\rangle\}$ subspace remain the same as the discussion in Sec.~\ref {sec:ParametricDrive}. (2) The residual effect from the counter-rotating termsis on the swapping between $\vert G \rangle$ and $\vert E \rangle$ with suppressed strength $\frac{J}{\Delta}\Omega_{\Phi}$.  Furthermore, the resonant parametric induced transition frequency between state  $\{\vert D\rangle, \vert B\rangle\}$ is around $2J/2\pi\cong 1.4 $ GHz, which is far away from the transition frequency $2\Delta/2\pi \cong 10 $ GHz between $\{\vert G\rangle, \vert E\rangle\}$. Therefore, it is safe to activate one channel without affecting the other. As a result, despite the strong inter-qubit coupling $J$, the Jaynes-Cummings approach well describes the main features of our study.
\parskip 0.1in

\section{BEYOND ROTATING WAVE APPROXIMATION} 
\label{sec:RWA}
\begin{figure} [b]
\includegraphics[width=80mm]{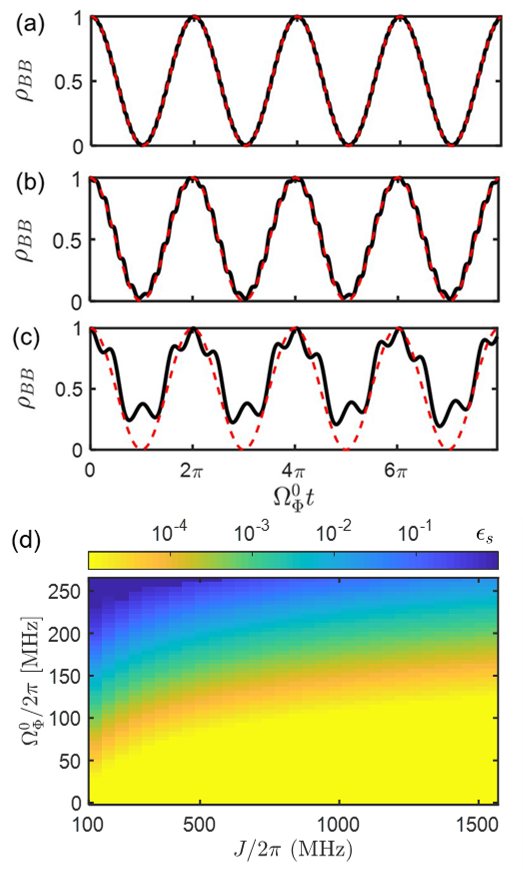}
\caption{Examination of the rotating wave approximation. The initial condition $\rho(t=0)=\vert B \rangle \langle B \vert$, bias point $\bar{\Phi}=0.25\Phi_0$ and the driving frequency at zero detuning $\Delta_{\Phi} = 0$ are set in the simulation.(a)(b)(c) illustrates the snapshots of $\rho_{BB}(t)$(black solid lines) at different $2J/\Omega^0_{\Phi}$. The dashed red lines are the sinusoidal fitting. The parametric driving amplitude is fixed at $\Omega^0_{\Phi} /2\pi = 23.3 $ MHz and the inter-qubit coupling $J$ is varied. 
(a) $2J/\Omega^0_{\Phi} = 15.8$, the resulting $\epsilon_s =2.6\times10^{-4} $. 
(b) $2J/\Omega^0_{\Phi} = 5.8$, $\epsilon_s = 2.1\times10^{-3} $. 
(c) $2J/\Omega^0_{\Phi} = 2.1$, $\epsilon_s = 4.6\times10^{-2} $.
(d) The sinusoidal likelihood $\epsilon_s$(colored) as a function of $\Omega^0_{\Phi}$ and inter-qubit coupling $J$.
}
\label{fig:RWA}
\end{figure} 
The RWA fails as $\Omega^0_{\Phi} \ll 2J$ no longer holds. Therefore, the design of inter-qubit coupling $J$ limits the maximum available $\Omega^0_{\Phi}$. The examination of RWA is evaluated by the correlation deviation factor $\epsilon_s$, defined as

\begin{equation}
\epsilon_s(\rho_{BB},f) = 1- \frac{\sum_t{(\rho_{BB}(t)-\bar{\rho}_{BB}) (f(t)-\bar{f}) }}{\sqrt{\sum{{(\rho_{BB}(t)-\bar{\rho}_{BB})}^2} \sum{{(f(t)-\bar{f})}^2}  }}.
\label{eq:corr}
\end{equation}

\noindent  It is the deviation from unity of the correlation of $\rho_{BB}(t)$, with the fitted sinusoidal function $f(t) = \sin{(\Omega^0_{\Phi} t + \phi)} + \frac{1}{2}$. Here $\rho(t=0) = \vert B \rangle \langle B \vert$ and the decoherence is nulled for simplicity.

Fig.~\ref{fig:RWA} illustrates the response to the parametric driving protocol beyond RWA. The RWA works well as the level spacing $J/2\pi > 100$ MHz for $\Omega^0_{\Phi}/2\pi <100$ MHz ($\epsilon<10^{-2}$), giving large flexibility for the device design.

\section{FREQUENCY TUNABILITY} 
\label{sec:tunability}

\begin{figure}
\includegraphics[width=85mm]{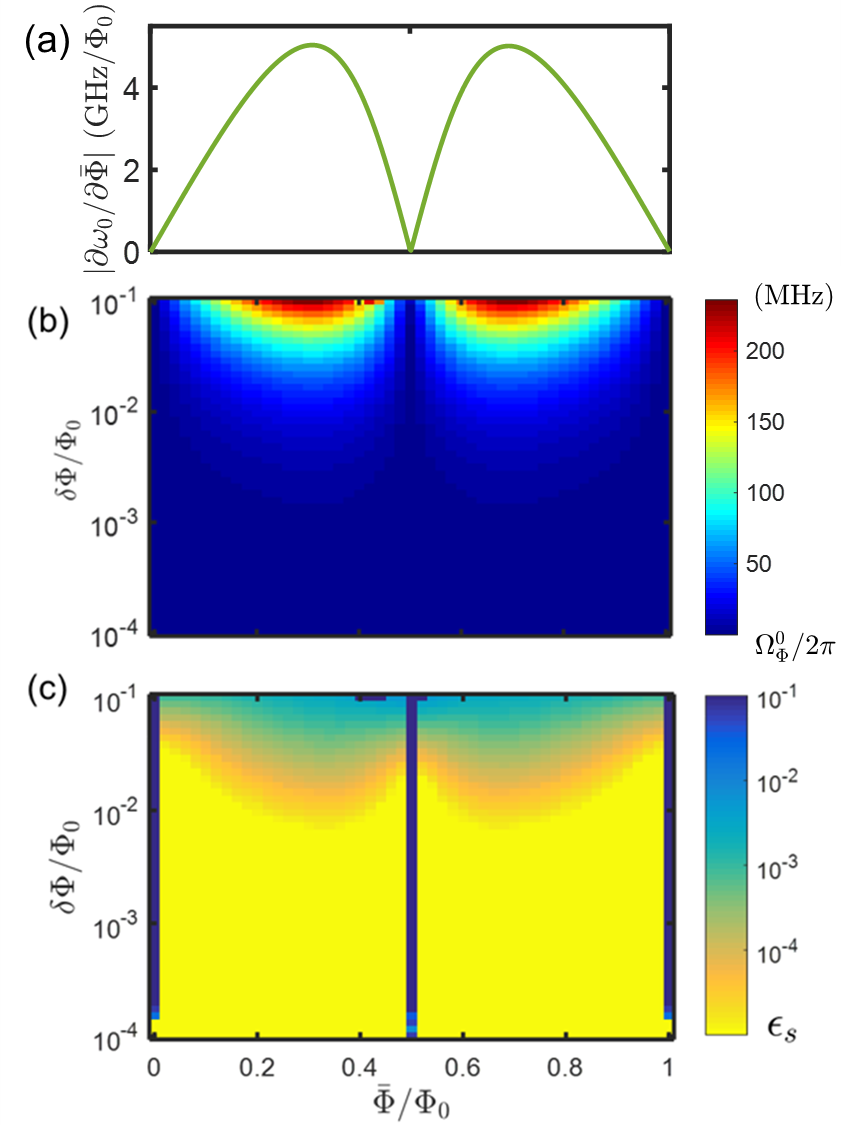}
\caption{Search over the tunability of the parametric drive with $J/2\pi = 700$ MHz. (a) Flux tunability $\partial\omega_0/\partial\Phi$ as a function of bias point $\bar{\Phi}$. (b) $\Omega^{\rm{fit}}_{\Phi}$ (colored) and (c) The RWA deviation factor $\epsilon_s$(colored) as a function of $\bar{\Phi}$ and modulation amplitude $\delta\Phi$.
}

\label{fig:Nonlin}
\end{figure} 

In Fig.~\ref{fig:Nonlin}, full mapping of the available $\Omega^0_{\Phi}$ at different bias point $\bar{\Phi}$ is studied. It indicates the operation window of the tunable $\Lambda$ system. The available $\Omega^0_{\Phi}$ is suppressed near the arc ($\bar{\Phi} = 0$) due to low flux sensitivity and the increased nonlinearity. Nevertheless, $\Omega^0_{\Phi}/2\pi \cong 10 $ MHz is still  available near the arc. It implies the EIT effect can be preserved with tunable transparency window over 1GHz. 

The unfavorable operation points are multiples of $\frac{1}{2}\Phi_0$ since the response of level modulation are the second harmonics.  It is reported that the modulation at the arc is doable~\cite{Didier18, Caldwell18, Reagor18} and the reduction of flux noise is achieved ~\cite{Didier19}. However, since the large nonlinearity, operation at the arc could slightly change $\omega_a$, and therefore break the degeneracy of the TCQ. While the resonance is retained as the two-qubit detuning  $\vert\omega_a - \omega_b\vert \ll J$, we leave that for further studies.
}


\begin{thebibliography}{0}
\bibitem{LZS}
In principle, Landau-Zener-St\"{u}ckelberg (LZS) type sidebands ($\pm k\Omega_{\Phi}, k\in$ integer) emerges around $\omega_{B}$ and $\omega_{D}$ as parametric drive is applied~\cite{Shevchenko10, Li13, Silveri15, Wen20}. The mean excitation of the $k$th band is proportional to $J_k(\Omega^0_{\Phi}/\omega_{\Phi})$, where $J_k$ denotes the $k$th-order Bessel function. However, as $\omega_{\Phi} \gg \Omega^0_{\Phi}$, the excitations of the $k\neq 0$ sidebands vanish and only the central frequencies $\omega_{B,D}$ survive.
\end{thebibliography}
\end{document}